\documentstyle[12pt,graphicx,epsf]{article}
\begin{document}

\def\AEF{A.E. Faraggi}
\def\NPB#1#2#3{{Nucl.\ Phys.}\/ {\bf B#1} (#2) #3}
\def\PLB#1#2#3{{Phys.\ Lett.}\/ {\bf B#1} (#2) #3}
\def\PRD#1#2#3{{Phys.\ Rev.}\/ {\bf D#1} (#2) #3}
\def\PRL#1#2#3{{Phys.\ Rev.\ Lett.}\/ {\bf #1} (#2) #3}
\def\PRT#1#2#3{{Phys.\ Rep.}\/ {\bf#1} (#2) #3}
\def\MODA#1#2#3{{Mod.\ Phys.\ Lett.}\/ {\bf A#1} (#2) #3}
\def\IJMP#1#2#3{{Int.\ J.\ Mod.\ Phys.}\/ {\bf A#1} (#2) #3}
\def\nuvc#1#2#3{{Nuovo Cimento}\/ {\bf #1A} (#2) #3}
\def\RPP#1#2#3{{Rept.\ Prog.\ Phys.}\/ {\bf #1} (#2) #3}
\def\APJ#1#2#3{{Astrophys.\ J.}\/ {\bf #1} (#2) #3}
\def\APP#1#2#3{{Astropart.\ Phys.}\/ {\bf #1} (#2) #3}
\def\etal{{\it et al\/}}

\newcommand{\nn}{\nonumber}
\newcommand{\Frac}[2]{\frac{\displaystyle{#1}}{\displaystyle{#2}}}
\newcommand{\bev}{\begin{verbatim}}
\newcommand{\beq}{\begin{equation}}
\newcommand{\beqa}{\begin{eqnarray}}
\newcommand{\beqn}{\begin{eqnarray}}
\newcommand{\eeqn}{\end{eqnarray}}
\newcommand{\eeqa}{\end{eqnarray}}
\newcommand{\eeq}{\end{equation}}
\newcommand{\Eev}{\end{verbatim}}
\newcommand{\bec}{\begin{center}}
\newcommand{\eec}{\end{center}}
\def\ie{{\it i.e.}}
\def\eg{{\it e.g.}}
\def\half{{\textstyle{1\over 2}}}
\def\nicefrac#1#2{\hbox{${#1\over #2}$}}
\def\third{{\textstyle {1\over3}}}
\def\quarter{{\textstyle {1\over4}}}
\def\m{{\tt -}}
\def\mass{M_{l^+ l^-}}
\def\p{{\tt +}}

\def\slash#1{#1\hskip-6pt/\hskip6pt}
\def\slk{\slash{k}}
\def\GeV{\;{\rm GeV}}
\def\TeV{\;{\rm TeV}}
\def\y{\;{\rm y}}

\def\l{\langle}
\def\r{\rangle}
\newcommand{\lsim}   {\mathrel{\mathop{\kern 0pt \rlap
  {\raise.2ex\hbox{$<$}}}
  \lower.9ex\hbox{\kern-.190em $\sim$}}}
\newcommand{\gsim}   {\mathrel{\mathop{\kern 0pt \rlap
  {\raise.2ex\hbox{$>$}}}
  \lower.9ex\hbox{\kern-.190em $\sim$}}}
\renewcommand{\thefootnote}{\fnsymbol{footnote}}
\setcounter{footnote}{0}

\begin{titlepage}
\samepage{
\setcounter{page}{1}

\rightline{OUTP-02-04P}
\rightline{March 2002}
\vspace{0.5cm}
\begin{center}
 {\Large \bf CP violation in realistic String Models\\  
                with family universal anomalous $U(1)$ \\}
\vspace{0.5cm}
 {\large Alon E. Faraggi\footnote{faraggi@thphys.ox.ac.uk}
 and          Oscar Vives \footnote{vives@thphys.ox.ac.uk}}\\
\vspace{.25cm}
{\it Theoretical Physics Department,\\
University of Oxford, Oxford, OX1 3NP, United Kingdom\\}
\end{center}
\begin{abstract}
The characteristic property of the $Z_2\times Z_2$ orbifold
compactification is the cyclic permutation symmetry between the three
twisted sectors. We discuss how this property, which is retained in a
class of realistic free fermionic string models, may be instrumental
in allowing fermion mass hierarchy while ensuring sfermion mass
degeneracy, irrespective of the dominant source of supersymmetry
breaking.  The cyclic symmetry is reflected in some models in the
existence of a family universal anomalous $U(1)_A$.  We analyze the
FCNC and $CP$ violation effects in a model with a dominant $U(1)_A$
SUSY breaking.  In this theories, new sources of FC are always
suppressed with respect to the average sfermion masses. We make a
phenomenological analysis of these effects and find that in most of
the cases they are in qualitative agreement with the phenomenological
limits.  The most sensitive low-energy observables to these new FC
sources are the $\varepsilon_K$ parameter, measuring indirect $CP$
violation in neutral kaon mixing and EDMs. These observables set
important constraints on the structure of the sfermion mass matrices,
but do not require a large fine-tuning of the initial parameters and
can be satisfied in most realistic constructions.

\end{abstract}
\smallskip}
\end{titlepage}

\section{Introduction}

The flavour puzzle in supersymmetric extensions of the Standard Model poses 
an especially interesting problem. On the one hand, there is a clear need
for flavour dependence to explain the fermion mass hierarchy. 
On the other hand, the absence of Flavour Changing Neutral Currents
(FCNC) at an observable rate suggests the need for flavour independent
symmetries, which force sfermion mass degeneracy. In addition one must
insure that non--trivial phases which may appear in the supersymmetry breaking
sector, do not produce Electric Dipole Moments (EDMs) that violate the 
experimental bounds \cite{EDM,abel}. Furthermore, as 
emphasized most recently in ref. \cite{abel}, even if the sfermion
masses are degenerate and the supersymmetric phases are suppressed,
non--universality in the soft trilinear A--terms may result in large
EDMs, merely due to the existence of CP violation in the 
Cabbibo--Kobayashi-Maskawa (CKM) mixing matrix. In ref. \cite{abel}
this was referred to as the ``String CP problem'',
to emphasize the fact that in string
theories one generically expects non--universal trilinear terms and
must necessarily provide for non-vanishing phases in the Yukawa matrices
\cite{abel}.

In the context of supersymmetric field theory models the problem is of course 
evaded by simply imposing that the relevant parameters have the needed
properties to avoid conflict with the experimental data. The riddle however
becomes more intricate in the context of superstring theories
in which the soft SUSY breaking parameters are, in general, expected
to be non--universal \cite{ibanezlust}. The problem is further exacerbated
due to our ignorance of the precise mechanism responsible for supersymmetry
breaking. Moreover, the SUSY breaking mechanism is, in general, expected
to involve nonperturbative dynamics, on which we have very little
calculational handle. Therefore, even if one scenario can offer
some remedy to the superstring flavour and CP problems, there is
no preference for any particular scenario, nor any reason for such a
remedy to survive the nonperturbative dynamics. 
These arguments suggest that what is needed is to seek
structures in string theory, or in particular string compactifications,
that are independent of our ignorance of the details of 
the SUSY breaking scenario and of the nonperturbative dynamics. 

In this paper we study the superstring flavour and CP problems in the context
of realistic string models with a family universal anomalous $U(1)$.
Supersymmetry breaking in this class of models was investigated
in refs. \cite{fh,pati}. The supersymmetry breaking mechanism
in these papers utilizes both an anomalous $U(1)$ gauge symmetry
and an effective mass term $m\sim1{\rm TeV}$ of some fields
in the massless string spectrum. It was shown that non--renormalizable
terms which contain hidden sector condensates, generate the required
suppression of the relevant mass term $m$ compared to the Planck
scale. While the non--vanishing $D$--term of the family
universal $U(1)_A$ led to squark degeneracy, those of
the family dependent $U(1)$s, remarkably enough, were found to
vanish for the solutions that were considered, owing to minimization of the
potential. This result therefore suggests the possibility that,
while the flavor $U(1)$ symmetries are responsible for generating
the fermion mass hierarchy, they do not necessarily spoil the sfermion
mass degeneracy \cite{pati}. Analysis of fermion mass textures
in the string models that we consider here was done in ref. \cite{fmasses}.
It was argued in ref. \cite{pati} that 
the phenomenological constraints motivate a combined $U(1)_A$--Dilaton
SUSY breaking scenario. It was then shown how the superstring FCNC problem, 
the analogous to the SUSY flavour problem in a superstring derived SUSY 
model, could be adequately resolved in this combined $U(1)_A$--Dilaton SUSY 
breaking scenario. 

In this work we extend the discussion of ref. \cite{pati}
and examine both the superstring flavour and CP problems in the context
of realistic string models with combined family universal
$U(1)_A$--Dilaton SUSY breaking scenario. Here we show that this
combined scenario can naturally evade the string CP problem
due to the suppression of the non--universal component
of the trilinear soft SUSY breaking parameters. 
We then discuss the general structure of the string models that 
produce the family universal anomalous $U(1)$. 
The root cause of the flavour universality
of $U(1)_A$ is the cyclic permutation symmetry that characterizes the
$Z_2\times Z_2$ orbifold compactification with standard
embedding \cite{fmasses,pati},
realized in the free fermionic models by the NAHE set \cite{nahe,z2z2}.
The cyclic permutation symmetry is the remarkable property
of the free fermionic models, or more precisely of the 
$Z_2\times Z_2$ orbifold, which may eventually prove to be instrumental
in resolving the string flavour and CP problems in any SUSY breaking
scenario. Namely, while in the present paper our focus
is on the family universality of the anomalous $U(1)_A$,
the cyclic permutation symmetry also exhibits itself in
the universality of the untwisted moduli, and in other sectors
of the models. It may therefore
prove to be instrumental to the understanding of the 
sfermion mass degeneracy in different SUSY breaking scenarios. 
In this respect the $U(1)_A$ SUSY breaking scenario that is analyzed
here and in ref. \cite{pati}, merely offers a glimpse
into the deeper structures that underly this class of
superstring compactifications. In section (\ref{stringin})
we discuss how the cyclic permutation symmetry may ensure 
sfermion universality irrespective of the dominant
source of supersymmetry breaking. 
Additionally, we discuss how the spectrum in the realistic free fermionic
models may divide in a way which allows fermion mass hierarchy 
while ensuring that the phenomenological constraints on the
sfermion mixing and phases are satisfied. 

\section{Anomalous $U(1)_A$ SUSY breaking}\label{ua1breaking}

In a large class of free fermionic string models with an anomalous $U(1)_A$ 
gauge symmetry, supersymmetry may be softly broken by the anomalous $D$-term.
At low energies these models yield a MSSM gauge group
with three generations and softly broken SUSY \cite{pati}.
Furthermore, the existence of free fermionic string models that
reproduce solely the MSSM states in the low energy, Standard Model charged, 
spectrum has also been demonstrated \cite{mssm}. 
These remarkable
properties render these string models as some of the most interesting 
constructions in string phenomenology. Issues like fermion mass textures
and CP violation in these models are discussed in refs. \cite{fmasses,mssm}.
In this letter we 
analyze the presence of FCNC and CP violation in these models. 
For this purpose we study two different realizations of free fermionic string 
models. We refer to them as model I \cite{alon1} and model II \cite{alon2}.

In model I, all scalar fields charged under $U(1)_A$ acquire a soft
mass given by, 
\beq [m_{\tilde{f}_i}^2]_{D_A} = g^2\left( Q_A^i \langle D_A
\rangle + Q_\chi^i \langle D_\chi \rangle \right).
\eeq
where $U(1)_\chi$ is a non-anomalous, family universal $U(1)$ that
gets a nonvanishing D-term in the model. In this case $\langle D_A \rangle 
= 2 \langle D_\chi \rangle = \sqrt{3/5} 
m^2/g^2$, and the charges are all family universal, although they vary for 
different members of the same family. Using the charges in \cite{alon1}
we obtain,
\beq
m^2_{Q_L}=m^2_{d_R}=3 m^2_{u_R}= 3 m^2_{L}= 3 m^2_{e_R}= m^2/4,
\label{u1} 
\eeq 
where $m$ leads to SUSY breaking
through the superpotential mass term
\beq
W \propto m \Phi \bar{\Phi} + \dots
\eeq 
In Ref. \cite{pati} this effective mass term in the superpotential 
corresponds to a $m \Phi_{45}\bar{\Phi}_{45}$ term, with 
$m \sim {\cal{O}}(1 \mbox{ TeV}) << M_{Pl}$ arising from higher dimensional
operators and hidden sector matter condensates.
Here all the MSSM fields have family universal charges 
under $U(1)_A$ and $U(1)_\chi$, and hence all of them receive an equal mass 
from the D-term. However, it is clear that the D-terms cannot contribute to 
gaugino masses and trilinear terms. Hence, in this scenario,
these soft terms can only appear
through higher dimensional operators. 
Gaugino masses are generated \cite{anomalous} from, 
\beq \lambda_{g} \int d^2 \theta \Frac{\Phi
\bar{\Phi}}{M_{Pl}^2} W_a W_a \longrightarrow \lambda_{g}
\Frac{\langle F_{\bar{\Phi}} \rangle \langle\Phi
\rangle}{M_{Pl}^2} =\lambda_g m \Frac{\sqrt{15}\xi}{5 M_{Pl}^2} \equiv 
\lambda_{g} \epsilon m/5 
\eeq 
with $\xi$ the Fayet-Iliopoulos term, that in this
model is equal to,
\beq 
\xi = \Frac{g^2 (\mbox{Tr} Q_A)}{192 \pi^2}
M_{Pl}^2 = \Frac{15 g^2} {16 \sqrt{15} \pi^2} M_{Pl}^2,
\eeq 
and
therefore we can estimate $\epsilon=0.095 \times g^2$. As $g^2$ is the 
unified coupling at the string scale and $\alpha_{GUT}=g^2/4 \pi \simeq 1/25$,
we have $\epsilon \simeq 1/20$. 
Similarly trilinear couplings come from superpotential terms with their 
flavour structure determined by a flavour symmetry as discussed in section 4
and therefore sufficiently similar to the Yukawa matrices. Still, we take
these flavour structures as free in the following,
\beq
W =\lambda^\prime Q H_u u^c \Frac{\Phi
\bar{\Phi}}{M_{Pl}^2} \longrightarrow \lambda^\prime
\Frac{\langle F_{\bar{\Phi}} \rangle \langle\Phi
\rangle}{M_{Pl}^2} = \lambda^\prime \epsilon m/5.
\eeq
Here the couplings $\lambda_g$, $\lambda^\prime$... respect the
string symmetries and may be suppressed and/or arise at 1 loop level.
For a scalar mass $m = 1$TeV, this implies that the gluino
mass at the electroweak scale
is $m_{\tilde{g}}(M_Z) \simeq\ 2.8\ m_\lambda(M_{\rm GUT}) = 
\ 30 \lambda{\rm GeV}$.
Clearly too small for the observed experimental limits. Hence we need
an additional contribution to the gaugino masses. To obtain an adequate
gaugino mass we can consider a combined anomalous
$U(1)$-Dilaton scenario. Here, scalar masses would get a dominant
contribution from $U(1)_A$ while the dilaton would generate
the main contribution to gaugino masses. 
These dilaton contributions give rise also to sfermion
masses and trilinear couplings although these contributions 
are family universal \cite{ibanez},
\beq
m_{1/2}=\pm \sqrt{3} m_{3/2}^{S}, \ \ \ \ \ m^2(\tilde{f}_i) = 
(m_{3/2}^{S})^2,
 \ \ \ \ \ A_{i j k}= - m_{1/2}
\label{dilaton}
\eeq
in terms of the gravitino mass, $m_{3/2}^{S}$, that arises from the
dilaton $F$--term. Notice that from here, the dilaton contributions to the
trilinear terms and to the gaugino masses have exactly the same phase.
Therefore, taking into account that in our scheme, dilaton provides always 
the great bulk of the gaugino mass, dilaton contribution to trilinear 
couplings play no role in $CP$ violation.   
Moreover, it is important to remember that flavour universal contributions
to the soft breaking terms, even with the two new SUSY phases 
(the $\mu$ and $A$ phases), can only contribute sizeably in $CP$ violation
in EDMs and $b\to s \gamma$, but never in $CP$ 
violation in kaon mixing ($\varepsilon_K$ and 
$\varepsilon^\prime/\varepsilon$) or in the $B^0$ $CP$
asymmetries \cite{flavourblind,wien}.

Still, we must take into account that sfermion masses may also receive
nonuniversal contributions from the K\"ahler potential \cite{anomalous},
\beq
(m^2_{\tilde{f}})_{K}\simeq \lambda \Frac{|\langle F_{\bar{\Phi}}\rangle 
|^2}{M^2_{Pl}}  \simeq \Frac{ \lambda m^2 \sqrt{15} \xi}{5 M^2_{Pl}} 
=  \lambda m^2 \epsilon/5
\label{lambdacouplings}
\eeq
Nevertheless, due to the $\epsilon$ suppression, these nonuniversal
contributions are always small and, as we will see in the next section,
generate reasonably small FCNC and $CP$ violation effects.

Similarly in model II the three generation sfermion masses are,
\beq
[m_{\tilde{f}_i}^2]_{D_A} = g^2\left( Q_A^i \langle D_A \rangle \right) 
= m^2/4,
\eeq 
where in this case $Q_A^i$ is not only family universal, but also 
intra--family universal, with $Q_A^i=1/\sqrt{12}$,
and $\langle D_A \rangle = \sqrt{3}/2\ m^2/g^2$.
Trilinear terms are obtained via higher order terms similar to model I,
\beq W =\lambda^\prime Q H_u u^c \Frac{\Phi
\bar{\Phi}}{M_{Pl}^2} \longrightarrow \lambda^\prime
\Frac{\langle F_{\bar{\Phi}} \rangle \langle\Phi
\rangle}{M_{Pl}^2} = \lambda^\prime \epsilon^\prime m/2, \eeq
and in this case $\epsilon^\prime = \sqrt{3}\xi/ M_{Pl}^2 = g^2 3/(8 \pi^2) 
\simeq 1/50$. Remarkably $\epsilon^\prime/2=\epsilon/5$ and the relative
suppression of trilinear terms with respect to $m$ is exactly the same as
in model I. Nonuniversal sfermion masses from the K\"ahler potential are 
now,
\beq
(m^2_{\tilde{f}})_{K}\simeq \lambda \Frac{|\langle F_{\bar{\Phi}}\rangle 
|^2}{M^2_{Pl}}  \simeq \Frac{ \lambda m^2 \sqrt{3} \xi}{2 M^2_{Pl}} 
=  \lambda m^2 \epsilon^\prime/2,
\eeq
and again the relative suppression with respect to $m$ is equal to model I.
This means that, at least in the down quark sector, the analysis of
FCNC and $CP$ violation will be analogous in both model I and II, as 
$\epsilon/5=\epsilon^\prime/2$. 

\section{$CP$ violation and FCNC}

As we have seen in the previous section, the free fermionic string models
of Refs. \cite{alon1,alon2} have a clear hierarchy between diagonal and 
off-diagonal elements,
\beq
\label{hierarchy}
\left[ m^2(\tilde{f}_i) \simeq \Frac{m^2}{4} + m^2_S \right] > 
\left[ (m^2_{\tilde{f}})_{i\neq j} \simeq \lambda_{i j} \Frac{m^2}{5} 
\epsilon \right]
\eeq
with $m^2_S$ the universal dilaton contribution to scalar masses.
This implies that, in general, the SUSY flavour problem as well as some 
aspects of the SUSY $CP$ problem are largely reduced in these scenarios. 
However, even with this suppression of off-diagonal elements, the 
FCNC and $CP$ violation observables are still extremely sensitive to 
some entries in the sfermion mass matrices. In the following, we analyze 
the possible contributions to FCNC and $CP$ violation observables in the 
quark sector.

In first place, we assume that the structure of the sfermion mass
matrices at the Planck scale is generically given in
Eq.~(\ref{hierarchy}).  The off--diagonal elements are roughly a factor
$\epsilon$ smaller than the diagonal ones. The next step is to use
the MSSM Renormalization Group Equations (RGE) to evolve the soft
breaking parameters from the Planck scale to the electroweak scale. In
fact, for simplicity, we identify the Planck scale and the GUT scale
and use the MSSM RGE up to a scale of $2 \times 10^{16}$ GeV. In this
RGE evolution the main effects are those associated with the strong
coupling, the top quark Yukawa coupling, and possibly the bottom quark
and tau lepton Yukawa couplings in the large $\tan \beta$ regime. We
can take the basis where up--quarks Yukawa couplings are diagonal,
$v_2\, Y_u = M_u$, and the off-diagonality in the down--quarks mass
matrix is simply given by the usual CKM mixing matrix, $v_1\, Y_d
=K_{CKM}^\dagger\cdot M_d$. Under these conditions, it is clear that we
can neglect small effects associated with Yukawa elements other than
$Y_{t t}$ and $Y_{b b}$. It is easy to see from the general MSSM RGE equations
\cite{RGE} that this implies that off-diagonal elements in the doublet
or singlet sfermion mass matrices are basically unchanged, while the
diagonal elements receive a flavour universal contribution except for a
small difference in the third generation masses (see for instance
Tables I and IV in \cite{wien}),
\beq
m^2_{\widetilde{D}_{(L,R)_i}} (M_W) \simeq 
6 \cdot m_{1/2}^2 + m_{\widetilde{D}_{(L,R)_i}}^2
\label{RGE}
\eeq 
{}From this point of view it is clear that these effects reduce the FCNC an 
$CP$ violation problems in the model. For instance, taking both the gaugino 
mass and the $U(1)_A$ contribution to sfermion masses roughly of the same 
order at the GUT scale, $m_{1/2} \simeq m /2$, and replacing in 
Eq. (\ref{RGE}) the different contributions from Eqs. (\ref{u1}) and 
(\ref{dilaton}), the average sfermion mass, $m^2_{\tilde{q}}$, at $M_W$ 
is given by $7.3\cdot m_{1/2}^2 \simeq 7.3 \cdot m^2 /4$. At the same time, 
as explained above,
off-diagonal terms are not largely modified, $(m^2_{\tilde{f}})_{i\neq j} 
\simeq \lambda_{i j} m^2\epsilon/5$. Hence the low energy FC 
effects are reduced nearly an order of magnitude by the above RGE factor. 
Nevertheless, we must keep in mind that in this model the sfermion and 
gaugino masses are in principle unrelated because they come from different
sources. On the other hand, the experimental constraints on sfermion and 
gaugino masses set for both a common lower limit of roughly $100$ GeV at 
the GUT scale. 
 
The RG evolution of the trilinear couplings is also similarly dominated 
by gluino contributions and the third generation Yukawa couplings.
Therefore, in the basis of diagonal up quark Yukawa couplings,
we have again that diagonal elements receive a 
large gaugino contribution at $M_W$, while the $(t,t)$ (and possibly $(b,b)$) 
element of the 
trilinear coupling, $Y^A_{ij}=A_{ij} Y_{ij}$, is reduced due to the top 
(and bottom) Yukawa coupling (see Table V in \cite{wien}). Once again 
off-diagonal 
elements in this basis are basically unchanged. In particular, the values of 
$A_{b b}$ depend on the value of $\tan \beta$. For instance, for low 
$\tan \beta \simeq 5$ we have,
\beq
A_{t} (M_W) \simeq 0.24 \cdot A_t^0 - 2 \cdot m_{1/2} \ \ \ \ \ \ \ 
A_{b} (M_W) \simeq 1 \cdot A_t^0 - 3.4 \cdot m_{1/2},  
\label{ARGE}
\eeq 
while for $\tan \beta \simeq 30$,
\beq
A_{t} (M_W) \simeq 0.25 \cdot A_t^0 - 2 \cdot m_{1/2} \ \ \ \ \ \ \ 
A_{b} (M_W) \simeq 0.74 \cdot A_t^0 - 2.9 \cdot m_{1/2},  
\label{ARGE30}
\eeq  
whereas, the flavour diagonal elements for the first two generations do not 
change strongly with $\tan \beta$,
\beq
A_{u} (M_W) \simeq 0.6 \cdot A_u^0 - 2.9 \cdot m_{1/2} \ \ \ \ \ \ \ 
A_{d} (M_W) \simeq 1 \cdot A_d^0 - 3.6 \cdot m_{1/2}  
\label{ARGE1}
\eeq  
The main feature of these RG evolution is the alignment of the trilinear
couplings and the gaugino masses. This alignment of the phases 
has important effects in $CP$ violation observables. 

To explore FCNC and $CP$-violation effects in the absence of a
completely defined flavour structure as in this case, it is convenient
to use the so-called Mass Insertion (MI) approximation
\cite{Hall-raby,MI}.  This approximation is defined in the SCKM basis
where fermion and sfermion matrices are rotated in parallel to the
basis where fermion masses are diagonal, such that neutral gaugino
couplings are flavour diagonal and the flavour change is produced by
non-diagonal sfermion propagators.  These propagators can be expanded
as a series in terms of, 
\beq (\delta_A)_{ij} =
(m^2_A)_{ij}/m^2_{\tilde{q}}, 
\eeq 
with $A=L,R,LR$ corresponding to the
left--handed sfermion mass matrix, right--handed sfermion mass matrix, or
in the left--right mixing sfermion mass matrices, respectively.  These
$\delta$ parameters are the so called Mass Insertions, and low energy
observables place constraints on the allowed size of these MI 
\cite{MI,MIupdate,pokorski}.

First, we analyze the $L$ and $R$ mass insertions.
Here we take the off-diagonal element as $(m^2_{\tilde{q}})_{K}$ in Eq. 
(\ref{hierarchy}),
and this implies, 
\beq (\delta_{L,R})_{ij}=\Frac{\lambda_{ij} m^2 \epsilon/5}{7.3 a (m^2/4)} 
\simeq \Frac{6 \lambda_{ij} \epsilon}{55 a }= \Frac{\lambda_{ij}}{a}\ 
5.5 \times 10^{-3}, 
\label{MI}
\eeq 
for an average squark mass, $m^2_{\tilde{q}}= 7.3 \cdot (m^2/4) a$. In
principle we make no further assumptions and take $\lambda_{ij}/a
\simeq 1$. Therefore we can see that in general we expect these
off-diagonal contributions to be quite small. In fact, if we compare
with the phenomenological bounds in \cite{MI,MIupdate,pokorski}, we can see
that the only places where these contributions could generate large
effects at low energy is in the kaon sector and possibly in rare
leptonic decays as $\mu \to e \gamma$ \cite{nonuni,annual}. On the
contrary, in the $B$ sector, where a large signal of $CP$ violation
has been recently measured in the $B$ factories \cite{bfactories}, the
expected SUSY contributions are small. The experimental limits from
the $B$ mass difference are \cite{MI,MIupdate,becirevic},
\beqn
\label{limitsB}
\sqrt{\left|\mbox{Re} \left(\delta^{d}_{L} \right)_{13}^{2} \right|},
\sqrt{\left|\mbox{Re} \left(\delta^{d}_{R} \right)_{13}^{2} \right|}
\leq 9.8 \times 10^{-2}\\ 
\sqrt{\left|\mbox{Re} \left\{\left(\delta^{d}_{L}
\right)_{13}\left(\delta^{d}_{R} \right)_{13}\right\}\right|}\leq 1.8 \times
10^{-2}\nn 
\eeqn 
with an average squark mass equal to the gluino mass at the
electroweak scale and equal to $500$ GeV. Notice that these limits
vary with the ratio $m^2_{\tilde{g}}/m^2_{\tilde{q}}$ and scale with
$m^2_{\tilde{q}}/(500 \mbox{ GeV})^2$.  Comparing the expected MI in
this model, Eq.~(\ref{MI}), and the limits in Eq.~(\ref{limitsB}), we
can see that in the case where off-diagonality is only present in the
$L$ or in the $R$ mass matrix the maximum possible contribution in our
model can never give a sizeable contribution.  In principle, the
simultaneous presence of maximum $\delta_L$ and a $\delta_R$ could
still be observable. Moreover, we have to take into account that due
to the fact that the $CP$ violating phase in $B$--$\bar{B}$ mixing is
${\cal{O}}(1)$, these limits are also approximately valid for the
imaginary part. In the presence of $L$ and $R$ MI simultaneously, it
is still possible that these contributions may be observed in the $CP$
asymmetries of the $B$ system. Nevertheless, in general, we would
expect that not all the off-diagonal entries are of the maximum size
in Eq.~(\ref{MI}). For instance, SUSY contributions would be smaller
if $\lambda/a <1$ or the average sfermion masses are somewhat
larger. Therefore, it is difficult to envision a sizeable contribution
in the $B$ system.  Another observable in the $B$ system is the decay
$b \to s \gamma$ but it is only sensitive to $LR$ MI and we discuss it
below.

More interesting are the contributions in the kaon sector. The kaon mass
difference as can be seen from the updated bounds in \cite{MIupdate}
can still receive a large contribution in the simultaneous presence
of $L$ and $R$ MI. This is due to the fact that in this case, most of
the kaon mass difference is already given in the SM. This fact already
constrains the model, although the experimental value of $\varepsilon_K$ 
gives rise to a much stronger constraint \cite{MI,MIupdate,pokorski}.
As generically we expect both the real and imaginary parts of these MI
of the same order we consider only the $\varepsilon_K$ bound,  
\beqn
\label{limits} 
\sqrt{\left|\mbox{Im} \left(\delta^{d}_{L} \right)_{12}^{2} \right|},
\sqrt{\left|\mbox{Im} \left(\delta^{d}_{R} \right)_{12}^{2} \right|}
\leq 6.1 \times 10^{-3}\\ 
\sqrt{\left|\mbox{Im} \left\{ \left(\delta^{d}_{L}
\right)_{12}\left(\delta^{d}_{R} \right)_{12}\right\}\right|}\leq 1.3 \times
10^{-4}\nn
\eeqn 
Once again, comparing these bounds with the naive estimate in Eq. (\ref{MI}), 
we can see here that a single MI in the right-- or left--handed 
sector should be in qualitative agreement with the phenomenological limits. 
In fact, in a complete theory, we would also expect some relation among
the flavour structures in the Yukawa couplings and the soft breaking masses.
Then, in the basis of diagonal Yukawa couplings, off--diagonal entries in the
sfermion mass matrices would be further suppressed with respect to the 
diagonal ones. Another necessary ingredient in this case is the presence of a 
non-vanishing $CP$ violating phase. Therefore the constraints on these matrix 
elements could be easier to satisfy if we assume that phases are
${\cal{O}}(0.1)$, although there is no special reason for this.
However, the simultaneous presence of large right and left 
MI can still be problematic for kaon phenomenology. The phenomenological 
limit in Eq (\ref{limits}) would imply a constraint in the off-diagonal
terms in the sfermion mass matrix,
\beq
\Frac{\sqrt{\mbox{Im} \left\{\lambda^{d_L}_{1 2}\lambda^{d_R}_{1 2}\right\}}}
{a} = \Frac{\sqrt{|\lambda^{d_L}_{1 2}| |\lambda^{d_R}_{1 2}|} 
\sin \alpha}
{a} \leq \Frac{1.3 \times 10^{-4}}{ 5.5 \times 10^{-3}} \simeq 0.024
\eeq 
with $\alpha=(\alpha_L+\alpha_R)/2$, the $CP$ phase of the 
$\lambda$--couplings in Eq. (\ref{lambdacouplings}), in the SCKM basis.
Hence, even in these models with suppressed nonuniversal contributions to
the sfermion mass matrices, the phenomenological constraint from
$\varepsilon_K$ has to be taken into account and restricts the allowed
flavour structure from the K\"ahler potential. 
However, we want to emphasize that here, these constraints do not
require a large fine-tuning of the initial parameters and could be
satisfied in realistic constructions.  
In the next section we will discuss these problems in the framework of a 
realistic string model.

A second source of flavour change comes from the nonuniversality in the
trilinear soft terms. These contributions give rise to chirality changing
MI, $(\delta_{LR})_{ij}$. In this case the off-diagonal entries are more 
difficult to estimate because the physical trilinear couplings are related
to the Yukawa matrices $Y^A_{ij}=A_{ij} Y_{ij}$ and this definition
is strongly dependent on the Yukawa basis. Indeed, in many flavour models,
this relation implies 
a strong suppression with light quark masses in the phenomenologically
interesting transitions. In fact, in some string or 
supergravity inspired models, although non-universal, it
has been argued that these trilinear
terms can be written as \cite{tatsuo}, $A_{ij}= A_i + A_j$, and then
the trilinear couplings are factorizable in matrix form,
\beq
Y^A_{ij}= \mbox{Diag}\left(A_1^L,A_2^L,A_3^L\right) \cdot Y + 
Y \cdot\mbox{Diag}\left(A_1^R,A_2^R,A_3^R\right)
\label{factor}
\eeq 
If the trilinear couplings possess this structure, and taking into account 
that off--diagonal elements are not largely affected by RGE evolution, we 
can estimate the $LR$ off--diagonal mass insertions \cite{nonuni},
\beqn
(\delta_{LR})_{i \neq j} = \Frac{1}{m^2_{\tilde{q}}} m_j ((A_2 - A_1) K_{i 2}
K_{j2}^* + (A_3 - A_1) K_{i 3} K_{j 3}^*) \simeq \\
\Frac{m_j \lambda^\prime m \epsilon/5}{7.3 a (m^2/4)} K_{\ell 2} K_{j\ell}^* 
\simeq \Frac{ \lambda^\prime}{a}\ \Frac{m_j}{500 \mbox{ GeV}}\ 
6.7 \times 10^{-3},   \nn
\eeqn  
where $m_j$ is the mass of the heaviest quark involved in the coupling and
$K_{i j}$ the matrix that diagonalizes the Yukawa couplings in this
basis\footnote{It is important to keep in mind that these
off-diagonal MI do not 
depend on $\tan \beta$ and are directly proportional to fermion masses. 
Only the flavour diagonal MI have a contribution proportional to $m_i \mu 
\tan \beta$.}.
Once more, to obtain the order of magnitude of this coupling we can assume
$\lambda^\prime/a \simeq 1$. These chirality changing MI give rise
to large flavour changing and $CP$ violation effects in several low energy
observables. In particular, the most important observables are the
$b \to s \gamma$ transition\footnote{This decay is specially constraining
in the large $\tan \beta$ regime in any MSSM, even in the absence of new 
flavour sources in the soft breaking terms \cite{bsg}.} and the direct $CP$ 
violation in the kaon sector,
$\varepsilon^\prime/\varepsilon$ \cite{murayama,KKV}. 

In the case of the $b \to s \gamma$ transition, the quark mass involved would
be $m_b$ which implies a further suppression of $6 \times 10^{-3}$.
Hence, we have,
\beq
(\delta^d_{LR})_{2 3 }\simeq \Frac{ \lambda^\prime}{a}\ 
4 \times 10^{-5} 
\eeq 
while the phenomenological limit is,
\beqn
\left|\left(\delta^{d}_{LR} \right)_{23} \right| \leq 1.6 \times 10^{-2}.
\eeqn
Therefore, it is clear that these $LR$ MI do not have observable effects in 
the $b \to s \gamma$ transition.

The second observable which is sensitive to $LR$ MI is
$\varepsilon^\prime/\varepsilon$. In this case, if we have a factorizable 
structure in the trilinear terms, the fermion mass involved is $m_s$, and 
the MI is,
\beq
(\delta^d_{LR})_{1 2}\simeq \Frac{ \lambda^\prime}{a}\ 
2 \times 10^{-6} 
\eeq 
to be compared with the limit,
\beqn 
\mbox{Im}\left(\delta^{d}_{LR} \right)_{12} \leq 2.0 \times 10^{-5}
\eeqn
Therefore, in this case, due to the suppression of trilinear terms, no large 
contributions are possible. 
More generally, even in the case where the structure of
the trilinear terms is not factorizable we always have at least
a suppression associated with the highest mass in the down sector, i.e. 
$m_b/(500 \mbox{ GeV})$. In that case, we would have a MI,
\beq
(\delta^d_{LR})_{1 2}\simeq \Frac{ \lambda^\prime}{a}\ 
K_{13}^d  K_{23}^{d\,*} \  4 \times 10^{-5} 
\eeq
with $K_{i j}^d$ the matrix that diagonalizes the down Yukawa matrix.
It is clear that the maximum possible value is $K_{13}^d  K_{23}^{d\,*} = 
1/2$, and in most reasonable models we would expect a much lower mixing,
for instance of the order of CKM mixings $\sim \lambda_C^5 = 3.2 
\times 10^{-4}$. Therefore, it seems highly improbable to have a sizeable 
contribution to this observable. This has to be compared with the situation
in general nonuniversal models where this observable can easily receive a 
sizeable contribution \cite{murayama,KKV}.

Next, we examine the contributions to electric dipole moments (EDMs).
In supersymmetric theories
the EDMs get new contributions at 1--loop from the $\mu$ and 
$A$ phases. In fact, since the dawn of the SUSY phenomenology era, it is 
well known that the experimental limits on the neutron EDM constrain 
$\phi_\mu$ and $\phi_A$ at $M_W$ to be roughly $\leq 10^{-2}$, unless sfermion
masses are pushed above ${\cal{O}}(1)$TeV. However, as we have seen in
Eqs.(\ref{ARGE}--\ref{ARGE1}), if we evolve these constraints to the GUT 
scale, the bounds on the initial phase of the $A$ terms are largely reduced 
and finally are $\phi_A\leq 10^{-1}$.
On the other hand, at 1--loop, the phase of the $\mu$ term\footnote{Always 
in the basis where we take the gaugino masses as real}
is invariant under the RGEs and only through the $B$ phase evolution some
scale dependence is introduced at $M_W$ in the basis where $B \mu$ is real. 
These effects are small for suppressed A-terms. Unfortunately, the mechanism 
to generate a $\mu$ term of the order of the electroweak scale, as 
phenomenology
requires, is strongly model dependent. In the following, we simply
assume that the relative phase between gaugino masses and the $\mu$ term 
is $\phi_\mu =0$. Therefore, we concentrate on the effects of trilinear
and Yukawa phases \cite{KKV,abel}.

Recently, it has been pointed out \cite{abel} that in string models
where non-universal trilinear terms are usually expected, even for 
purely real trilinear terms, you can expect large SUSY contributions
to EDMs simply from the phases in the Yukawa matrix. These effects were used 
to show the tight constraints on non-universality from EDM experiments
\cite{abel}. Moreover, when additional flavour
structures in the soft breaking terms are present, new phases become
observable and can give rise to large effects here, as they do in neutral kaon
mixing \cite{piai}.  However, although these contributions must be taken into
account in a generic model, they are not necessarily important in any
non-universal MSSM. For instance, if the trilinear terms are
factorizable, as in Eq. (\ref{factor}), or the Yukawa and trilinear matrices
hermitian \cite{bailin}, this problem is not present. Moreover, in a 
realistic string model, as the one we are
analyzing here, we show that the problem is also softened and in fact
absent in most reasonable constructions.

{}From here on, we consider the effects of an ${\cal{O}}(1)$ trilinear 
phase. Clearly this case includes also possible effects of Yukawa phases
\cite{abel}. As we discussed above
the $A$-terms in $U(1)_A$--Dilaton SUSY breaking
models are suppressed
relative to diagonal sfermion masses.
First, we  analyze the case of a factorizable structure in the trilinear 
couplings. Here, the contribution to the diagonal $LR$ mass insertion is 
\footnote{assuming that the $K$ matrices that diagonalize the fermion masses
have dominant terms $K_{i i} \simeq 1$ 
in the basis of diagonal soft masses},
\beqn
\label{LR} 
(\delta^q_{LR})_{i i} = \Frac{A_i m_i}{m^2_{\tilde{q}}}  \simeq 
\Frac{m_i \lambda^\prime m \epsilon/5}{7.3 a (m^2/4)} \simeq 
\Frac{\lambda^\prime}{a}
\ \Frac{m_i}{500 \mbox{ GeV}}\ 7.4 \times 10^{-3}, \\
(\delta^l_{LR})_{i i} = \Frac{A_i m_i}{m^2_{\tilde{l}}}  \simeq 
\Frac{m_i \lambda^\prime m \epsilon/5}{1.7 a (m^2/4)} \simeq 
\Frac{\lambda^\prime}{a}
\ \Frac{m_i}{100 \mbox{ GeV}}\ 1.5 \times 10^{-2}. \nn
\eeqn  
Again, if we are interested in the electron or neutron EDMs, the additional
suppression from the light quark mass is enough to satisfy the experimental
limits \cite{MI,Hgedm} \footnote{Here we consider only the bounds on the
quark electric dipole moments. However the quark chromoelectric dipole 
moments from Hg atomic experiments can be more restrictive \cite{Hgedm}
and give rise to stronger constraints.},
\beqn
&\left|\mbox{Im}\left(\delta^{d}_{LR} \right)_{11} \right| \leq 3.0 \times
10^{-6} ,\ \ \
\left|\mbox{Im}\left(\delta^{u}_{LR} \right)_{11} \right| \leq 5.9 \times
10^{-6},&\\
&\left|\mbox{Im}\left(\delta^{l}_{LR} \right)_{11} \right| \leq 3.7 \times
10^{-7}.&\nn
\eeqn
Here limits in the squark sector assume an average sfermion mass of $500$
GeV, while in the slepton sector an average mass of $100$ GeV is assumed.
The mass suppression is $m_u/ 500 \mbox{ GeV} \simeq 1 \times 10^{-5}$, 
$m_d/500 \mbox{ GeV} \simeq 2 \times 10^{-5}$ and 
$m_e/100 \mbox{ GeV} \simeq 5 \times 10^{-6}$. Therefore all the EDM bounds 
are easily satisfied in this case.

A third interesting observable is the muon EDM. Although the current 
experimental limits, $d_\mu < 1.05
\times 10^{-18}$ ecm, cannot provide a constraint on the $LR$ MI, 
a recent proposal has been made at BNL for a dedicated experiment
to reach a sensitivity of $10^{-24}$ ecm \cite{BNL}.
If we take this value as the future experimental limit this implies
that the future limit on the MI will be,
\beqn
&\left|\mbox{Im}\left(\delta^{l}_{LR} \right)_{22} \right| \leq 5.3
\times 10^{-5},
\eeqn
and again the mass suppression is $m_\mu/100 \mbox{ GeV} \simeq 1 \times 
10^{-3}$, which gives a value roughly a factor 5 below the expected future 
bounds.   

Even in the case where this factorizable structure is not present, it is
clear that in the leptonic sector, or in the down quark sector, the minimal
suppression is given by, $m_\tau/(100 \mbox{ GeV}) = 1.7 \times 10^{-2}$,
$m_b/(500 \mbox{ GeV}) = 6 \times 10^{-3}$. However, there is little
suppression
for the top quark in the up sector, $m_t/(500 \mbox{ GeV}) = 0.35$. In any 
case, it is clear that this minimal suppression is not enough even including 
the suppression of $\epsilon/5$. In these
cases we would need to take also into account the mixings of the third
generation with the first and second generations,
\beqn
(\delta^q_{LR})_{i i} = |K_{i3}|^2\Frac{c_3 A m_3}{m^2_{\tilde{q}}}  \simeq 
|K_{i3}|^2 \Frac{c_3 m_3 \lambda^\prime m \epsilon/5}{7.3 a (m^2/4)} 
\simeq  \Frac{\lambda^\prime}{a}
\ \Frac{c_3|K_{i3}|^2 m_3}{500 \mbox{ GeV}}\ 7.4 \times 10^{-3}  \\
(\delta^l_{LR})_{i i} = |K_{i3}|^2\Frac{c_3 A m_3}{m^2_{\tilde{l}}}  \simeq 
|K_{i3}|^2 \Frac{c_3 m_3 \lambda^\prime m \epsilon/5}{1.7 a (m^2/4)} 
\simeq  \Frac{\lambda^\prime}{a}
\ \Frac{c_3|K_{i3}|^2 m_3}{100 \mbox{ GeV}}\ 1.5 \times 10^{-2}  
\eeqn
with $c_i$ a coefficient taking into account the RGE effects on the 
third generation trilinear couplings, Eqs. (\ref{ARGE}--\ref{ARGE1}). We
take $c_t \simeq 0.25$ and $c_b \simeq c_\tau \simeq 1$.  
Therefore we would have a maximum contribution,
\beqn
(\delta^d_{LR})_{1 1} \simeq  \Frac{\lambda^\prime}{a}
\ |K_{13}^d|^2\ 4.4 \times 10^{-5}, \\
(\delta^u_{LR})_{1 1} \simeq  \Frac{\lambda^\prime}{a}
\ |K_{13}^u|^2\ 6.5 \times 10^{-4},   
\eeqn
for squarks of $500$ GeV. In the case of sleptons, we have,
\beqn
(\delta^e_{LR})_{1 1} \simeq (\delta^e_{LR})_{2 2} \simeq  
\Frac{\lambda^\prime}{a}\ |K_{i3}^l|^2\ 2.5 \times 10^{-4} 
\eeqn
with an average slepton mass of $100$ GeV.

Clearly, it is very difficult to make a definite statement without a complete
theory of flavour that provides the relative rotation matrices between
quarks and squarks, $K$. At least a very reasonable assumption 
\cite{nonuni,annual,piai} is to take these matrices to be of the same order as
the CKM mixing matrix. This was indeed the choice in \cite{abel}.
In that case, $K_{13}^u \simeq K_{13}^d \simeq 10^{-2}$, and obviously
the bounds are in this case easily satisfied due to the 
$\epsilon$ suppression of the trilinear terms in anomalous $U(1)$ models.
In the leptonic sector, things are not that straightforward because of our
ignorance of the leptonic mixings\footnote{Notice that neutrino mixings
are largely influenced by the Seesaw mechanism, see for instance 
\cite{king,buchmuller,masina}.}. In this case, the limits on the electron
EDM (and future limits on the muon EDM) would require a mixing matrix roughly
of the same order as the CKM matrix, as found for instance in
\cite{king,buchmuller} even with maximal neutrino mixings.

We conclude that EDM constraints are not violated by nonuniversality of
the A-terms in most 
reasonable scenarios in $U(1)_A$--Dilaton SUSY breaking models.

\section{String insights}\label{stringin}

In this section
we discuss how the cyclic permutation symmetry may ensure 
sfermion universality irrespective of the dominant
source of supersymmetry breaking. 
Additionally, we discuss how the spectrum in the realistic free fermionic
models may divide in a way which allows fermion mass hierarchy 
while ensuring that the phenomenological constraints on the
sfermion mixings and phases are satisfied. 
As we have seen in the previous section, even with the suppression of the
nonuniversal contributions in anomalous $U(1)$ SUSY breaking models,
there are some 
low energy observables, such as the EDMs and $\varepsilon_K$, that place
further constraints on the flavour structure of the nonuniversal
soft terms.
In first instance, the nonuniversality in the trilinear
terms may contribute to EDMs.
The important property of family universal $U(1)_A$ SUSY breaking
is the suppression of the trilinear $A$--terms, which
have the general form 
\beq
A_{\alpha\beta\gamma}=F^m\left[{\hat K}_m+
\partial_m\log Y_{\alpha\beta\gamma}-
\partial_m\log({\tilde K}_\alpha{\tilde K}_\beta{\tilde K}_\gamma)\right].
\label{aterms}
\eeq
Here, Latin indices refer to the hidden sector fields, 
while Greek indices refer to the observable fields;
the K\"ahler potential is expanded in observable fields
as $K={\hat K}+{\tilde K}_\alpha\vert C^\alpha\vert^2+ \cdots$,
and ${\hat K}_m=\partial_m{\hat K}$. The sum in $m$ runs over
all of the SUSY breaking fields. Combined with a non--vanishing
dilaton $F$--term this SUSY breaking scenario produces viable
gaugino and sfermion masses, while maintaining adequate family
universality, which is required by the FCNC and
CP phenomenological constraints.

While the combined $U(1)_A$--Dilaton SUSY breaking scenario
is phenomenologically appealing, 
the difficulty lies in the fact that we do not have 
a dynamical reason to prefer this scenario over other 
SUSY breaking sources. The natural question to ask
is what are the underlying string structures
that yield the family universal $U(1)_A$ and whether
this structures may also be preserved in other SUSY
breaking scenarios. In this perspective the anomalous
$U(1)_A$ charges merely provide us with a window
to the properties of the underlying string compactification, 
which resulted in family universality \cite{sano}. In turn, we may 
expect these highlighted characteristics to be preserved
in other sectors of the models, as well as in the nonperturbative
regimes. Namely, in the class of models that we discuss,
the family universality originates in the most robust
structure of the underlying string compactification. 

The class of models under consideration are the 
free fermionic heterotic--string models.
The construction of these models has been amply discussed
in the past as well as numerous phenomenological studies. 
We refer the interested reader to the original literature
for the details \cite{reviews}. Here we focus on the properties
of the models that 
pertain to sfermion universality and the string CP problem.

The free fermionic models are specified in terms
of a set of boundary condition basis vectors. 
The class of models that we discuss here are spanned 
by a set eight such basis vectors. 
The first five basis vectors $\{{\bf 1},S,b_1,b_2,b_3\}$,
in the models of interest here, consist of the so called NAHE set \cite{nahe},
which yields after the generalized GSO projections
an $N=1$ supersymmetric $SO(10)\times SO(6)^3\times E_8$
gauge group, with 48 generations in the 16 representation of $SO(10)$.
The remaining three boundary conditions basis vectors, 
typically denoted by $\{\alpha, \beta, \gamma\}$,
break the $SO(10)$ symmetry to one of its subgroups
and reduce the number of generations to three. 
One from each of the sectors $b_1$, $b_2$ and $b_3$. 
The importance of the NAHE set lies in its correspondence
with $Z_2\times Z_2$ orbifold compactification \cite{z2z2}.
The three sectors $b_1$, $b_2$ and $b_3$
correspond to the three twisted sectors of the $Z_2\times Z_2$
orbifold, whereas the basis vectors $\{\alpha, \beta, \gamma\}$
correspond to Wilson lines in an orbifold formalism.

The characteristic property of the $Z_2\times Z_2$ orbifold 
compactification, which is the origin for the emergence
of a family universal anomalous $U(1)$, is the cyclic 
permutation symmetry between the three sectors
$b_1$, $b_2$ and $b_3$ with respect to their
left-- and right--moving world--sheet charges.
In general, the boundary condition basis
vectors $\{\alpha,\beta,\gamma\}$
break the permutation symmetry between the
three light generations. However, in string models
of refs. \cite{alon1} and \cite{alon2} the permutation
symmetry is maintained, which is the reason for the
existence of a family universal anomalous $U(1)$ in
these models. It is important to note that this universality 
structure of the $Z_2\times Z_2$ orbifold compactification
is also reflected in other sectors of the models, in particular
for the untwisted moduli. Thus, even if SUSY breaking is dominated
by the untwisted moduli sector, squark degeneracy is still expected.

As correctly emphasized in ref. \cite{abel} the difficulty 
in understanding the flavour and string CP problems lies in the simultaneous
requirement to generate fermion mass hierarchy and sfermion mass
degeneracy. From Eq. (\ref{aterms}) it is seen that in order 
not to generate non--universal $A$--terms the fields that
generate the Yukawa matrices should have a vanishing $F$--term
and hence should not break supersymmetry. To understand how this
situation may come about in the string models we recall how
the fermion mass hierarchy is generated in these models. 
The most relevant feature for our purposes is the identification
of the sectors and states that contribute to the generation of
the fermion mass hierarchy and those that do not. Thus, it is those
sectors and states of the second kind, namely those that do not
participate in the generation of the fermion mass hierarchy,
that may play the role in the SUSY breaking dynamics.

In addition to the three chiral generations arising from the twisted
sectors $b_1$, $b_2$ and $b_3$ the untwisted sector of the models
produces three pairs of Higgs doublets, $\{h_i,{\bar h}_i\}$ $(i=1,2,3)$.
Typically the free fermionic
string models contain one additional sector that produces electroweak
Higgs doublets $\{h_{\alpha\beta},{\bar h}_{\alpha\beta}\}$.
This sector arises from a combination of the $\alpha$
and $\beta$ basis vector and is denoted as the $\alpha\beta$--sector.
The sectors $b_j+2\gamma$ give rise to states in the $16_j$ representation
of the hidden $SO(16)_H$ gauge group. 
These are decomposed under the final unbroken
$SO(16)_H$ subgroup, which typically
contains two unbroken non--Abelian gauge groups. Being, for example, 
$SU(3)_H$ and $SU(5)_H$ in the models of refs. \cite{alon1, alon2}.

The fermion mass terms arise from $N^{th}$--order
superpotential terms of the form 
$cgf_if_jh\phi^{^{N-3}}$ or
$cgf_if_j{\bar h}\phi^{^{N-3}}$, where $c$ is a 
calculable coefficient, $g$ is the gauge coupling at the unification 
scale,  $f_i$, $f_j$ are the fermions from
the sectors $b_1$, $b_2$ and $b_3$, $h$ and ${\bar h}$ are the light 
Higgs doublets, and $\phi^{N-3}$ represent a product
of Standard Model singlet fields
that get a VEV and produce a suppression factor
${({{\langle\phi\rangle}/{M}})^{^{N-3}}}$ relative to the cubic
level terms. Here $M\sim10^{18}{\rm GeV}$ is a scale
related to the heterotic--string unification scale. 
At the cubic level only the couplings
$\{u_jQ_j+N_jL_j\}{\bar h}_j$ and $\{d_jQ_j+e_jL_j\}h_j$
are allowed. Note that each generation couples to a different
Higgs pair, and that at this level the cyclic permutation symmetry
is retained. As the anomalous $U(1)$ Fayet--Iliopoulos term
breaks supersymmetry
near the Planck scale, we must assign VEVs to some Standard
Model singlets, along flat $F$ and $D$ directions. In this process
some of the nonrenormalizable terms become effective renormalizable
operators.
At the same time some of the Higgs doublet representations
receive large mass. For specific solutions only two Higgs doublets
remain massless down to the electroweak scale. In typical analysis
these have consisted of ${\bar h}_1$ and $h_{\alpha\beta}$. 
Analysis of the nonrenormalizable terms
up to order $N=8$ reveals the following structure \cite{fmasses},
\beq
{M_U\sim\left(\matrix{\epsilon,a,b\cr
                    {\tilde a},A,c \cr
                    {\tilde b},{\tilde c},\lambda_t\cr}\right);{\hskip .2cm}
M_D\sim\left(\matrix{\epsilon,d,e\cr
                    {\tilde d},B,f \cr
                    {\tilde e},{\tilde f},C\cr}\right);{\hskip .2cm}
M_E\sim\left(\matrix{\epsilon,g,h\cr
                    {\tilde g},D,i \cr
                    {\tilde h},{\tilde i},E\cr}\right)},
\label{fmatrices}
\eeq
where $\epsilon\sim({{\Lambda_{Z^\prime}}/{M}})^2\approx0$.
$U(1)_{Z^\prime}$ is the Abelian symmetry in $SO(10)$,
which is orthogonal to the Standard Model.
The diagonal terms in capital letters represent leading 
terms that are suppressed by singlet VEVs \cite{fmasses},
and $\lambda_t=O(1)$.
The mixing terms are generated by hidden sector states from the
sectors $b_j+2\gamma$ and are represented by small letters. They 
are proportional to $({{\langle{TT}\rangle}/{M}^2})$, 
where ${\langle{TT}\rangle}$ represents the
VEVs of these hidden sector matter states.
The states from the sector $b_3$ are identified with the
lightest generation \cite{fmasses}.

The important aspect for our purpose here is to identify in the string 
models the sectors and states that contribute to the products of
fields $\phi^n$ which induce the fermion mass hierarchies.
The analysis of the nonrenormalizable terms that contribute to
the fermion mass matrices was carried out in detail in ref. \cite{fmasses}.
Inspection of the nonrenormalizable terms in \cite{fmasses}
then reveals that fields that contribute to all the leading terms 
of the mass matrices in Eq. (\ref{fmatrices}) arise from the 
following sectors 
\beq
\{{\rm untwisted~sector}, \alpha\beta-{\rm sector},b_j+2\gamma\} ~~(j=1,2,3)
\label{sectors}
\eeq
The mixing terms have the general underlying form 
$16_i16_j1016_i16_j.$ The first two $16$s are
observable chiral $SO(10)$ representations
and the last two are hidden $SO(16)$ vectorial
representations. The $10$ are the vectorial $SO(10)$ representations that
produce the light Higgs multiplets.
The hidden $SO(16)$ gauge group
is typically broken to two non--Abelian group factors. 
Being, for example, $SU(5)$ and $SU(3)$ in the models
of refs. \cite{alon1,alon2}. The vectorial $16$ representation
is therefore broken into $5\oplus{\bar5}$ and $3\oplus{\bar3}$
of the hidden $SU(5)$ and $SU(3)$ group factors, respectively.
Typically the matter states of one of this group factors
are included in the products $\phi^n$ that generate the 
fermion mass terms, whereas the matter states under the
second group factor form matter condensates that
may trigger supersymmetry breaking \cite{fh}. 

A very robust and model independent solution to the flavour and string CP
problems is to require that the states that appear in the 
products $\phi^n$ that generate the fermion mass terms 
do not have a non--vanishing $F$--term.
In the class of models under study here
we note from the discussion above that
this simply means that none of the states that arise 
from the sectors in Eq. (\ref{sectors}) has a non--vanishing $F$--term.
This is a very modest requirement.
The fields that trigger supersymmetry breaking should 
arise from other sectors in the models. They may originate,
for example, from the Wilsonian $SO(10)$ breaking sectors
that contain combinations of the boundary condition basis
vectors $\{\alpha,\beta, \gamma\}$. As we note here
the states from these sectors do not appear in the
products $\phi^n$ that induce the fermion mass hierarchies
in (\ref{fmatrices}).

As discussed in section (\ref{ua1breaking}),
a further source of FCNC and $CP$ violation is the nonuniversality in the
sfermion masses that arises from the K\"ahler potential and affects
$\varepsilon_K$.   
We comment here on the possible role of the cyclic
permutation symmetry in suppressing non--universal
contributions from the K\"ahler potential. Correspondingly
to Eq. (\ref{aterms}) the K\"ahler potential contribution to the
sfermion masses is given by
$m_{ij}^2\sim F^aF_b\partial_a\partial^b K_{ij}$,
where $\partial_a$ denotes $\partial_a/\partial\Phi^a$.
Therefore, we may in general expect non--universal contributions
through the dependence
of the K\"ahler potential on the fields $\Phi^a$
with $F^a\ne0$. In the case of the mixed universal $U(1)_A$--Dilaton
SUSY breaking scenario, as discussed above, these non--universal
contributions are suppressed relative to the dominant
universal contributions. However, also in the case that
$F$--term of some field dominates we may expect that the
cyclic permutation symmetry might play a role in ensuring
sfermion degeneracy. Namely, we may expect that
the permutation symmetry is preserved in other
sectors of the models, and not merely in the $U(1)_A$ charges.
Examining the spectrum of the model of ref. \cite{alon1}
we note that additional sectors in the model obey the permutation symmetry.
for example, the sectors $b_i+b_j+\alpha+\beta\pm\gamma+(I)$, $i,j=1,2,3$
$i\ne j$ in table 2 of \cite{alon1}. This suggests a similar situation to
the 
the family universal anomalous $U(1)_A$. Namely, $U(1)_A$ is a combination
of three generation dependent world--sheet currents. Its universality arises 
precisely because it is such a combination. Similarly, due to
the permutation symmetry, we may envision the SUSY breaking
fields to be a combination of the states from the sectors  
$b_i+b_j+\alpha+\beta\pm\gamma+(I)$ that preserves the permutation
symmetry hence retain the universal contribution to the sfermion
masses\footnote{Nevertheless, it has been shown in Ref. \cite{gaillard} that
it may be difficult to generate $CP$ violation without a contribution from
the geometric moduli which generically would have non-vanishing F-terms}.

\section{Conclusions}

In this letter, we have analyzed the FCNC and $CP$ violation effects in a 
model with a dominant family universal anomalous $U(1)$ source of
Supersymmetry
breaking.
In this theories, new sources of FC are always suppressed with respect
to the average sfermion masses. We made a phenomenological analysis of
these effects and found that in most of the cases they are in qualitative 
agreement with the phenomenological limits. 
The most sensitive low-energy observables to these new FC sources are
the $\varepsilon_K$ parameter, measuring indirect $CP$ violation in the
neutral kaon mixing and EDMs. These observables set important constraints 
in the structure of the sfermion mass matrices.
Nevertheless, these constraints do not
require a large fine-tuning of the initial parameters and can be
satisfied in most realistic constructions, with some mild assumptions
about the flavour structure of the K\"ahler potential.  

Additionally, we discussed in this paper how the fermion mass
hierarchy may arise in the string models without inducing
non--universal $A$--terms.  This is achieved provided that the fields
which induce the fermion mass hierarchy do not break supersymmetry and
have a vanishing $F$--term.  From the structure of the free fermionic
models we noted that this requirement places a restriction on the
sectors from which the SUSY breaking fields may arise.

Finally, the family universality of the anomalous $U(1)$ in the free
fermionic models originates in the cyclic permutation symmetry of the
$Z_2\times Z_2$ orbifold compactification with respect to the twisted
sectors. In the three generation models under consideration this
cyclic permutation symmetry is retained with respect to the horizontal
$U(1)$ charges of the chiral generations. The cyclic permutation
symmetry between the twisted sectors is the characteristic property of
the $Z_2\times Z_2$ orbifold. As we discussed here, preservation of
the cyclic permutation symmetry is reflected in other sectors of the
models and may be instrumental to understand the sfermion mass
degeneracy irrespective of the dominant source of supersymmetry
breaking.  Naturally, a complete solution to the sfermion mass
degeneracy problem cannot be attained before a simultaneous
detailed analysis of
the fermion mass hierarchy is obtained. However, short of this
ambitious and still distant goal, we note how the underlying structure
of the realistic free fermionic models may be instrumental in allowing
fermion mass hierarchy while ensuring sfermion mass degeneracy.

\bigskip
\centerline{\bf Acknowledgments}
\smallskip

The work of A.F. is supported by PPARC. O.V.
acknowledges support from the EC under
contract HPRN-CT-2000-0148.

\newpage

\end{document}